# Computation of Optimal Control Problems with Terminal Constraint via Variation Evolution

Sheng ZHANG, Bo LIAO, and Fei LIAO

(2017.12)

*Abstract:* Enlightened from the inverse consideration of the stable continuous-time dynamics evolution, the Variation Evolving Method (VEM) analogizes the optimal solution to the equilibrium point of an infinite-dimensional dynamic system and solves it in an asymptotically evolving way. In this paper, the compact version of the VEM is further developed for the computation of Optimal Control Problems (OCPs) with terminal constraint. The corresponding Evolution Partial Differential Equation (EPDE), which describes the variation motion towards the optimal solution, is derived, and the costate-free optimality conditions are established. The explicit analytic expressions of the costates and the Lagrange multipliers adjoining the terminal constraint, related to the states and the control variables, are presented. With the semi-discrete method in the field of PDE numerical calculation, the EPDE is discretized as finite-dimensional Initial-value Problems (IVPs) to be solved, with common Ordinary Differential Equation (ODE) numerical integration methods.

## I. INTRODUCTION

Optimal control theory aims to determine the inputs to a dynamic system that optimize a specified performance index while satisfying constraints on the motion of the system. It is closely related to engineering and has been widely studied [1]. Because of the complexity, Optimal Control Problems (OCPs) are usually solved with numerical methods. Various numerical methods are developed and generally they are divided into two classes, namely, the direct methods and the indirect methods [2]. The direct methods discretize the control or/and state variables to obtain the Nonlinear Programming (NLP) problem, for example, the widely-used direct shooting method [3] and the classic collocation method [4]. These methods are easy to apply, whereas the results obtained are usually suboptimal [5], and the optimal may be infinitely approached. The indirect methods transform the OCP to a Boundary-value Problem (BVP) through the optimality conditions. Typical methods of this type include the well-known indirect shooting method [2] and the novel symplectic method [6]. Although be more precise, the indirect methods often suffer from the significant numerical difficulty due to the ill-conditioning of the Hamiltonian dynamics, that is, the stability of costates dynamics is adverse to that of the states dynamics [7]. The recent development, representatively the Pseudo-spectral (PS) method [8], blends the two types of methods, as it unifies the NLP and the BVP in a dualization view [9]. Such methods inherit the advantages of both types and blur their difference.

Theories in the control field often enlighten strategies for the optimal control computation, for example, the non-linear variable transformation to reduce the variables [10]. Recently, a new Variation Evolving Method (VEM), which is enlightened by the states evolution within the stable continuous-time dynamic system, is proposed for the optimal control computation [11][12][13]. The VEM also synthesizes the direct and indirect methods, but from a new standpoint. The Evolution Partial Differential Equation (EPDE), which describes the evolution of variables towards the optimal solution, is derived from the viewpoint of variation motion, and the optimality conditions will be asymptotically met under this frame. In Refs. [11] and [12], besides the states and the controls, the costates are also employed in developing the EPDE, and this increases the complexity of the computation. Ref. [13] proposed the compact version of the VEM that uses only the original variables, but it can only handles a class of OCPs with free terminal states. In this paper, the VEM is further developed to accommodate the OCPs with terminal constraint, and the corresponding evolution equations are derived.

Throughout the paper, our work is built upon the assumption that the solution for the optimization problem exists. We do not describe the existing conditions for the purpose of brevity. Relevant researches such as the Filippov-Cesari theorem are documented in [14]. In the following, first the principle of the VEM and the results regarding the OCPs without terminal constraint are reviewed. Then the VEM for OCPs with terminal constraint is developed. During this course, the equivalent costate-free optimality conditions are established, and the explicit analytic solution of costates and Lagrange multipliers in the classic treatment are obtained. Later illustrative examples are solved to verify the effectiveness of the method.

## II. PRELIMINARIES

### A. Principle of VEM

The VEM is a newly developed method for the optimal solutions. It originates from the dynamics stability theory in the control field.

**Lemma 1** [15] (with small adaptation): For a continuous-time autonomous dynamic system like

$$\dot{x} = f(x) \tag{1}$$

where $x \in \mathbb{R}^n$ is the state, $\dot{x} = \dfrac{dx}{dt}$ is its time derivative, and $f : \mathbb{R}^n \to \mathbb{R}^n$ is a vector function. Let $\hat{x}$, contained within the domain $\mathbb{D}$, be an equilibrium point that satisfies $f(\hat{x}) = 0$ and $\mathbb{D}$. If there exists a continuously differentiable function $V : \mathbb{D} \to \mathbb{R}$ such that

i) $V(\hat{x}) = c$ and $V(x) > c$ in $\mathbb{D}/\{\hat{x}\}$.

ii) $\dot{V}(x) \leq 0$ in $\mathbb{D}$ and $\dot{V}(x) < 0$ in $\mathbb{D}/\{\hat{x}\}$.

where $c$ is a constant. Then $x = \hat{x}$ is an asymptotically stable point in $\mathbb{D}$.

Lemma 1 aims to the dynamic system with finite-dimensional states, and it may be directly generalized to the infinite-dimensional case as

**Lemma 2**: For an infinite-dimensional dynamic system described by

$$\frac{\delta y(x)}{\delta t} = f(y, x) \tag{2}$$

or presented equivalently in the PDE form as

$$\frac{\partial y(x,t)}{\partial t} = f(y, x) \tag{3}$$

where "$\delta$" denotes the variation operator and "$\partial$" denotes the partial differential operator. $x \in \mathbb{R}$ is the independent variable, $y(x) \in \mathbb{R}^n(x)$ is the function vector of $x$, and $f : \mathbb{R}^n(x) \times \mathbb{R} \to \mathbb{R}^n(x)$ is a vector function. Let $\hat{y}(x)$, contained within a certain function set $\mathbb{D}(x)$, is an equilibrium function that satisfies $f(\hat{y}(x),x) = \mathbf{0}$. If there exists a continuously differentiable functional $V: \mathbb{D}(x) \to \mathbb{R}$ such that

i) $V(\hat{y}(x)) = c$ and $V(y(x)) > c$ in $\mathbb{D}(x)/\{\hat{y}(x)\}$.

ii) $\dot{V}(y(x)) \leq 0$ in $\mathbb{D}(x)$ and $\dot{V}(y(x)) < 0$ in $\mathbb{D}(x)/\{\hat{y}(x)\}$.

where $c$ is a constant. Then $y(x) = \hat{y}(x)$ is an asymptotically stable solution in $\mathbb{D}(x)$.

In the system dynamics theory, from the stable dynamics, we may construct a monotonously decreasing function (or functional) $V$, which will achieve its minimum when the equilibrium is reached. Inspired by it, now we consider its inverse problem, that is, from a performance index function to derive the dynamics that minimize this performance index, and optimization problems are just the right platform for practice. Under this thought, the optimal solution is analogized to the equilibrium of a dynamic system and is anticipated to be obtained in an asymptotically evolving way. Accordingly, a virtual dimension, the variation time $\tau$, is introduced to implement the idea that a variable $x(t)$ evolves to the optimal solution to minimize the performance index within the dynamics governed by the variation dynamic evolution equations. Fig. 1 illustrates the variation evolution process of the VEM in solving the OCP. Through the variation motion, the initial guess of variable will evolve to the optimal solution.

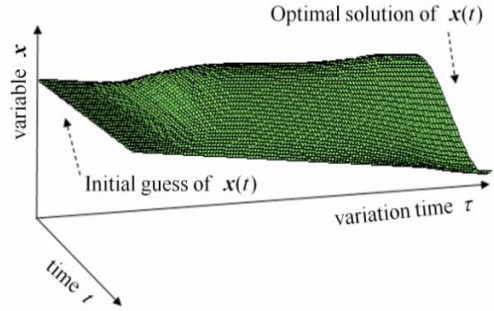

Fig. 1. The illustration of the variable evolving along the variation time $\tau$ in the VEM.

The VEM bred under this idea is demonstrated for the unconstrained calculus-of-variations problems and the OCPs with free terminal states [11][13]. The variation dynamic evolution equations, derived under the frame of the VEM, may be reformulated as the EPDE and the Evolution Differential Equation (EDE), by replacing the variation operator "$\delta$" with the partial differential operator "$\partial$" and differential operator "d". Since the right function of the EPDE only depends on the time $t$, it is suitable to be solved with the well-known semi-discrete method in the field of PDE numerical calculation [16]. Through the discretization along the normal time dimension, the EPDE is transformed to the finite-dimensional Initial-value Problem (IVP) to be solved, with the mature Ordinary Differential Equation (ODE) integration methods. Note that the resulting IVP is defined with respect to the variation time $\tau$, not the normal time $t$.

*B. Results for OCPs with free terminal states*

To clear the development, first the results for the OCPs with free terminal states, obtained under the frame of the VEM, is recalled.

**Problem 1**:Consider performance index of Bolza form

$$J = \varphi(\mathbf{x}(t_f), t_f) + \int_{t_0}^{t_f} L(\mathbf{x}(t), \mathbf{u}(t), t) \mathrm{d}t \tag{4}$$

subject to the dynamic equation

$$\dot{x} = f(x, u, t) \tag{5}$$

where $t \in \mathbb{R}$ is the time. $x \in \mathbb{R}^n$ is the state vector and its elements belong to $C^2[t_0, t_f]$. $u \in \mathbb{R}^m$ is the control vector and its elements belong to $C^1[t_0, t_f]$. The function $L: \mathbb{R}^n \times \mathbb{R}^m \times \mathbb{R} \to \mathbb{R}$ and its first-order partial derivatives are continuous with respect to $x$, $u$ and $t$. The function $\varphi: \mathbb{R}^m \times \mathbb{R} \to \mathbb{R}$ and its first-order and second-order partial derivatives are continuous with respect to $x$ and $t$. The vector function $f: \mathbb{R}^n \times \mathbb{R}^m \times \mathbb{R} \to \mathbb{R}^n$ and its first-order partial derivatives are continuous and Lipschitz in $x$, $u$ and $t$. The initial time $t_0$ is fixed and the terminal time $t_f$ is free. The initial boundary conditions are prescribed as

$$x(t_0) = x_0 \tag{6}$$

and the terminal states are free. Find the optimal solution $(\hat{x}, \hat{u})$ that minimizes $J$, i.e.

$$(\hat{x}, \hat{u}) = \arg\min(J) \tag{7}$$

For this class of OCPs, presuming we already have a feasible initial solution of the states and control variables that satisfies Eqs. (5) and (6), the variation dynamic evolution equations derived from the VEM are

$$\frac{\delta x}{\delta \tau} = \int_{t_0}^{t} \boldsymbol{\Phi}_o(t, s) f_u(s) \frac{\delta u}{\delta \tau}(s) \mathrm{d}s \tag{8}$$

$$\frac{\delta u}{\delta \tau} = -K p_u(t) \tag{9}$$

$$\frac{\delta t_f}{\delta \tau} = -k_{t_f} \left( L + \varphi_t + \varphi_x^{\mathrm{T}} f \right) \Big|_{t_f} \tag{10}$$

where

$$p_u(t) = L_u + f_u^{\mathrm{T}} \varphi_x + f_u^{\mathrm{T}} \left( \int_t^{t_f} \boldsymbol{\Phi}_o^{\mathrm{T}}(\sigma, t) \left( L_x(\sigma) + \varphi_{tx}(\sigma) + \varphi_{xx}^{\mathrm{T}}(\sigma) f(\sigma) + f_x(\sigma)^{\mathrm{T}} \varphi_x(\sigma) \right) \mathrm{d}\sigma \right) \tag{11}$$

$K$ is the $m \times m$ dimensional positive-definite matrix and $k_{t_f}$ is a positive constant. $\boldsymbol{\Phi}_o(t, s)$ is the $n \times n$ dimensional state transition matrix from time point $s$ to time point $t$, which satisfies

$$\frac{\partial}{\partial t} \boldsymbol{\Phi}_o(t, s) = f_x(t) \boldsymbol{\Phi}_o(t, s) \tag{12}$$

Use the partial differential operator "$\partial$" and the differential operator "d", the variation dynamic evolution equations may be reformulated to get the following EPDE and EDE as

$$\frac{\partial}{\partial \tau} \begin{bmatrix} x \\ u \end{bmatrix} = \begin{bmatrix} \int_{t_0}^{t} \boldsymbol{\Phi}_o(t, s) f_u(s) \frac{\partial u}{\partial \tau}(s) \mathrm{d}s \\ -K p_u \end{bmatrix} \tag{13}$$

$$\frac{\mathrm{d} t_f}{\mathrm{d} \tau} = -k_{t_f} \left( L + \varphi_t + \varphi_x^{\mathrm{T}} f \right) \Big|_{t_f} \tag{14}$$

with the definite conditions including the initial guess of $t_f$, i.e., $t_f \big|_{\tau=0} = \tilde{t}_f$, and,

$$\begin{bmatrix} x(t, \tau) \\ u(t, \tau) \end{bmatrix} \bigg|_{\tau=0} = \begin{bmatrix} \tilde{x}(t) \\ \tilde{u}(t) \end{bmatrix} \tag{15}$$

where $\tilde{x}(t)$ and $\tilde{u}(t)$ are the initial feasible solution. The solution at $\tau = \infty$, determined by Eqs. (13), (14) and (15), will satisfy

$$p_u = 0 \tag{16}$$

$$L(t_f) + \phi_t(t_f) + \varphi_x^{\mathrm{T}}(t_f) f(t_f) = 0 \tag{17}$$

Eqs. (16) and (17) are the first-order optimality conditions for Problem 1 without the employment of costates. They are proved equivalent to the traditional ones with costates [17].

### III. VEM FOR OCPs WITH TERMINAL CONSTRAINT

#### A. Problem definition

In this paper, we consider the OCPs with terminal constraint that is defined as

**Problem 2:** Consider performance index of Bolza form

$$J = \varphi(x(t_f), t_f) + \int_{t_0}^{t_f} L(x(t), u(t), t) \mathrm{d}t \tag{18}$$

subject to the dynamic equation

$$\dot{x} = f(x, u, t) \tag{19}$$

where $t \in \mathbb{R}$ is the time. $x \in \mathbb{R}^n$ is the state vector and its elements belong to $C^2[t_0, t_f]$. $u \in \mathbb{R}^m$ is the control vector and its elements belong to $C^1[t_0, t_f]$. The function $L: \mathbb{R}^n \times \mathbb{R}^m \times \mathbb{R} \to \mathbb{R}$ and its first-order partial derivatives are continuous with respect to $x$, $u$ and $t$. The function $\varphi: \mathbb{R}^m \times \mathbb{R} \to \mathbb{R}$ and its first-order and second-order partial derivatives are continuous with respect to $x$ and $t$. The vector function $f: \mathbb{R}^n \times \mathbb{R}^m \times \mathbb{R} \to \mathbb{R}^n$ and its first-order partial derivatives are continuous and Lipschitz in $x$, $u$ and $t$. The initial time $t_0$ is fixed and the terminal time $t_f$ is free. The initial and terminal boundary conditions are respectively prescribed as

$$x(t_0) = x_0 \tag{20}$$

$$g(x(t_f), t_f) = 0 \tag{21}$$

where $g: \mathbb{R}^n \times \mathbb{R} \to \mathbb{R}^q$ is a $q$ dimensional vector function with continuous first-order partial derivatives. Find the optimal solution $(\hat{x}, \hat{u})$ that minimizes $J$, i.e.

$$(\hat{x}, \hat{u}) = \arg\min(J) \tag{22}$$

Compared with the OCPs with free terminal states as defined in Problem 1, Problem 2 covers them because Problem 1 may be considered as a special case with $g = 0$.

#### B. Derivation of variation dynamic evolution equations

Instead of circumventing Problem 2 by constructing an equivalent unconstrained functional problem that has the same extremum [11], we will address Problem 2 in the way similar to Ref. [13]. We also consider the problem within the feasible solution domain $\mathbb{D}_o$, in which any solution satisfies Eqs. (19), (20) and (21). First we transform the Bolza performance index to the equivalent Lagrange type, i.e.

$$J = \int_{t_0}^{t_f} \left( \varphi_t + \varphi_x^{\mathrm{T}} f(x, u, t) + L(x, u, t) \right) \mathrm{d}t \tag{23}$$

where $\varphi_t$ and $\varphi_x$ are the partial derivatives notated as before. Differentiating Eq. (23) with respect to the variation time $\tau$ gives

$$\frac{\delta J}{\delta \tau} = (\varphi_t + \varphi_x^T f + L)\Big|_{t_f} \frac{\delta t_f}{\delta \tau} + \int_{t_0}^{t_f} \left( (\varphi_{tx}^T + f^T \varphi_{xx} + \varphi_x^T f_x + L_x^T) \frac{\delta x}{\delta \tau} + (\varphi_x^T f_u + L_u^T) \frac{\delta u}{\delta \tau} \right) dt \quad (24)$$

where $\varphi_{tx}$ and $\varphi_{xx}$ are second-order partial derivatives in the form of (column) vector and matrix, and $f_x$ and $f_u$ are the Jacobi matrixes. For the solutions in $\mathbb{D}_o$, $\frac{\delta x}{\delta \tau}$ and $\frac{\delta u}{\delta \tau}$ are related because of Eq. (19), and they need to satisfies the following variation equation as

$$\frac{\delta \dot{x}}{\delta \tau} = f_x \frac{\delta x}{\delta \tau} + f_u \frac{\delta u}{\delta \tau} \quad (25)$$

with the initial condition $\frac{\delta x}{\delta \tau}\Big|_{t_0} = 0$. Note that $f_x$ and $f_u$ are time-dependent matrixes linearized at the feasible solution $x(t)$ and $u(t)$. Eq. (25) is a linear time-varying equation and has a zero initial value. Thus according to the linear system theory [18], its solution may be explicitly expressed by Eq. (8). Use Eq. (8) and follow the same derivation as Ref. [13], we again obtain

$$\frac{\delta J}{\delta \tau} = (\varphi_t + \varphi_x^T f + L)\Big|_{t_f} \frac{\delta t_f}{\delta \tau} + \int_{t_0}^{t_f} p_u^T \frac{\delta u}{\delta \tau} dt \quad (26)$$

Different from the OCPs with free terminal states, we cannot use the variation dynamic evolution equations (9) and (10) to achieve $\frac{\delta J}{\delta \tau} \leq 0$. This is because in that way the terminal constraint (21) is not guaranteed, and we have to find a solution that not only guarantees $\frac{\delta J}{\delta \tau} \leq 0$ but also satisfies the variation equation of Eq. (21) as

$$g_{x_f} \int_{t_0}^{t_f} \Phi_o(t_f, t) f_u(t) \frac{\delta u}{\delta \tau}(t) dt + (g_{x_f} f + g_{t_f}) \frac{\delta t_f}{\delta \tau} = 0 \quad (27)$$

**Theorem 1**: The following variation dynamic evolution equations guarantees that the solution stays in the feasible domain and the performance index $\frac{\delta J}{\delta \tau} \leq 0$

$$\frac{\delta u}{\delta \tau} = -K\left(p_u + f_u^T \Phi^T(t_f, t) g_{x_f}^T \pi\right) \quad (28)$$

$$\frac{\delta t_f}{\delta \tau} = -k_{t_f}\left(L + \varphi_t + \varphi_x^T f + \pi^T(g_{x_f} f + g_{t_f})\right)\Big|_{t_f} \quad (29)$$

where $K$ is the $m \times m$ dimensional positive-definite matrix and $k_{t_f}$ is a positive constant, $p_u$ is defined in Eq. (11) and the parameter vector $\pi \in \mathbb{R}^q$ is the solution of the linear matrix equation

$$M\pi = -r \quad (30)$$

where the $q \times q$ dimensional matrix $M$ and the $q$ dimension vector $r$ are

$$M = g_{x_f}\left(\int_{t_0}^{t_f} \Phi(t_f, t) f_u K f_u^T \Phi^T(t_f, t) dt\right) g_{x_f}^T + k_{t_f}(g_{x_f} f + g_{t_f})(g_{x_f} f + g_{t_f})^T\Big|_{t_f} \quad (31)$$

$$r = g_{x_f}\left(\int_{t_0}^{t_f} \Phi(t_f, t) f_u K p_u dt\right) + k_{t_f}(g_{x_f} f + g_{t_f})(\varphi_t + \varphi_x^T f + L)\Big|_{t_f} \quad (32)$$

Moreover, under the evolution equations (28) and (29), $\frac{\delta J}{\delta \tau} = 0$ occurs only when

$$p_u + f_u^T \Phi^T(t_f,t) g_{x_f}^T \pi = 0 \tag{33}$$

$$(\varphi_t + \varphi_x^T f + L + \pi^T(g_{x_f} f + g_{t_f}))\big|_{t_f} = 0 \tag{34}$$

and the optimal value of $\pi$ satisfies

$$\begin{bmatrix} M_{s1} \\ M_{s2} \end{bmatrix} \pi = - \begin{bmatrix} r_{s1} \\ r_{s2} \end{bmatrix} \tag{35}$$

where the $q \times q$ dimensional matrixes $M_{s1}$, $M_{s2}$ and the $q$ dimension vectors $r_{s1}$, $r_{s2}$ are

$$M_{s1} = g_{x_f} \left( \int_{t_0}^{t_f} \Phi(t_f,t) f_u f_u^T \Phi^T(t_f,t) dt \right) g_{x_f}^T \tag{36}$$

$$M_{s2} = (g_{x_f} f + g_{t_f})(g_{x_f} f + g_{t_f})^T \big|_{t_f} \tag{37}$$

$$r_{s1} = g_{x_f} \int_{t_0}^{t_f} \Phi(t_f,t) f_u p_u \, dt \tag{38}$$

$$r_{s2} = (g_{x_f} f + g_{t_f})(\varphi_t + \varphi_x^T f + L)\big|_{t_f} \tag{39}$$

Proof: We will derive Eqs. (28) and (29) though the optimization theory. Reformulate Eq. (26) as a constrained optimization problem subject to Eq. (27) as

$$J_{t1} = (\varphi_t + \varphi_x^T f + L)\big|_{t_f} \frac{\delta t_f}{\delta \tau} + \int_{t_0}^{t_f} p_u^T \frac{\delta u}{\delta \tau} dt \tag{40}$$

Note that now $\frac{\delta u}{\delta \tau}$ is the decision variable and $\frac{\delta t_f}{\delta \tau}$ is the decision parameter. Since the minimum of this optimization problem may be negative infinity, to penalize too large decision variable (parameter), we introduce another performance index $J_{t2}$ to formulate a Multi-objective Optimization Problem (MOP) as

$$\begin{aligned} &\min(J_{t1}, J_{t2}) \\ &s.t. \\ &g_{x_f} \int_{t_0}^{t_f} \Phi_o(t_f,t) f_u(t) \frac{\delta u}{\delta \tau}(t) dt + (g_{x_f} f + g_{t_f}) \frac{\delta t_f}{\delta \tau} = 0 \end{aligned} \tag{41}$$

where

$$J_{t2} = \frac{1}{2k_{t_f}} (\frac{\delta t_f}{\delta \tau})^2 + \int_{t_0}^{t_f} \frac{1}{2} (\frac{\delta u}{\delta \tau})^T K^{-1} \frac{\delta u}{\delta \tau} dt \tag{42}$$

We use the weighting method to solve the Pareto optimal solution of this MOP, and the resulting performance index is

$$J_{t3} = a J_{t1} + b J_{t2} \tag{43}$$

where $a \geq 0$, $b \geq 0$ and $a + b = 1$. When $a = 1$, $b = 0$, we get a solution that minimizes $J_{t1}$. When $a = 0$, $b = 1$, we get a solution that minimize $J_{t2}$. Otherwise, we get a compromising solution. For this MOP, obviously in the case of $a = 0$, $b = 1$, the Pareto optimal solution is that $\frac{\delta t_f}{\delta \tau} = 0$ and $\frac{\delta u}{\delta \tau} = 0$, and now the value of performance indexes are $J_{t1} = 0$ and $J_{t2} = 0$. For any other cases, the compromising solution guarantees that $J_{t1} \leq 0$.

Set $a = \frac{1}{2}$, $b = \frac{1}{2}$ and introduce the Lagrange multiplier $\pi \in \mathbb{R}^q$ to adjoin the constraint, we may get the unconstrained optimization problem as

$$J_{t4} = \frac{1}{2}J_{t1} + \frac{1}{2}J_{t2} + \frac{1}{2}\pi^{\text{T}}\left(\boldsymbol{g}_{\boldsymbol{x}_f}\int_{t_0}^{t_f}\boldsymbol{\Phi}_o(t_f,t)\boldsymbol{f}_{\boldsymbol{u}}(t)\frac{\delta \boldsymbol{u}}{\delta \tau}(t)\mathrm{d}t + \boldsymbol{g}_{\boldsymbol{x}_f}\boldsymbol{f}\frac{\delta t_f}{\delta \tau} + \boldsymbol{g}_{t_f}\frac{\delta t_f}{\delta \tau}\right) \quad (44)$$

Use the fist-order optimality conditions, i.e., $\dfrac{\partial J_{t4}}{\partial \left(\dfrac{\delta \boldsymbol{u}}{\delta \tau}\right)} = \boldsymbol{0}$ and $\dfrac{\partial J_{t4}}{\partial \left(\dfrac{\delta t_f}{\delta \tau}\right)} = 0$, we may get Eqs. (28) and (29). Substitute Eqs. (28) and (29) into Eq. (27), we have

$$\boldsymbol{g}_{\boldsymbol{x}_f}\int_{t_0}^{t_f}\boldsymbol{\Phi}(t_f,t)\boldsymbol{f}_{\boldsymbol{u}}\boldsymbol{K}\left(\boldsymbol{p}_{\boldsymbol{u}} + \boldsymbol{f}_{\boldsymbol{u}}^{\text{T}}\boldsymbol{\Phi}^{\text{T}}(t_f,t)\boldsymbol{g}_{\boldsymbol{x}_f}^{\text{T}}\pi\right)\mathrm{d}t + k_{t_f}\left(\boldsymbol{g}_{\boldsymbol{x}_f}\boldsymbol{f} + \boldsymbol{g}_{t_f}\right)(\varphi_t + \varphi_{\boldsymbol{x}}^{\text{T}}\boldsymbol{f} + L + \pi_{q\times 1}^{\text{T}}\boldsymbol{g}_{\boldsymbol{x}_f}\boldsymbol{f})\Big|_{t_f} = \boldsymbol{0} \quad (45)$$

Further deduction gives

$$\left(\boldsymbol{g}_{\boldsymbol{x}_f}\left(\int_{t_0}^{t_f}\boldsymbol{\Phi}(t_f,t)\boldsymbol{f}_{\boldsymbol{u}}\boldsymbol{K}\boldsymbol{f}_{\boldsymbol{u}}^{\text{T}}\boldsymbol{\Phi}^{\text{T}}(t_f,t)\mathrm{d}t\right)\boldsymbol{g}_{\boldsymbol{x}_f}^{\text{T}} + k_{t_f}(\boldsymbol{g}_{\boldsymbol{x}_f}\boldsymbol{f} + \boldsymbol{g}_{t_f})(\boldsymbol{g}_{\boldsymbol{x}_f}\boldsymbol{f} + \boldsymbol{g}_{t_f})^{\text{T}}\Big|_{t_f}\right)\pi$$
$$= -\boldsymbol{g}_{\boldsymbol{x}_f}\left(\int_{t_0}^{t_f}\boldsymbol{\Phi}(t_f,t)\boldsymbol{f}_{\boldsymbol{u}}\boldsymbol{K}\boldsymbol{p}_{\boldsymbol{u}}\mathrm{d}t\right) - k_{t_f}(\boldsymbol{g}_{\boldsymbol{x}_f}\boldsymbol{f} + \boldsymbol{g}_{t_f})(\varphi_t + \varphi_{\boldsymbol{x}}^{\text{T}}\boldsymbol{f} + L)\Big|_{t_f} \quad (46)$$

Thus, with the definition of $\boldsymbol{M}$ in Eq. (31) and $\boldsymbol{r}$ in Eq. (32), Eq. (30) that determines $\pi$ is obtained.

Furthermore, Eq. (26) may be reformulated as

$$\frac{\delta J}{\delta \tau} = (\varphi_t + \varphi_{\boldsymbol{x}}^{\text{T}}\boldsymbol{f} + L)\Big|_{t_f}\frac{\delta t_f}{\delta \tau} + \int_{t_0}^{t_f}\boldsymbol{p}_{\boldsymbol{u}}^{\text{T}}\frac{\delta \boldsymbol{u}}{\delta \tau}\mathrm{d}t$$
$$= \left(\varphi_t + \varphi_{\boldsymbol{x}}^{\text{T}}\boldsymbol{f} + L + \pi^{\text{T}}(\boldsymbol{g}_{\boldsymbol{x}_f}\boldsymbol{f} + \boldsymbol{g}_{t_f})\right)\Big|_{t_f}\frac{\delta t_f}{\delta \tau} + \int_{t_0}^{t_f}\left(\boldsymbol{p}_{\boldsymbol{u}} + \boldsymbol{f}_{\boldsymbol{u}}^{\text{T}}\boldsymbol{\Phi}^{\text{T}}(t_f,t)\boldsymbol{g}_{\boldsymbol{x}_f}^{\text{T}}\pi\right)^{\text{T}}\frac{\delta \boldsymbol{u}}{\delta \tau}\mathrm{d}t \quad (47)$$
$$-\pi^{\text{T}}\left(\boldsymbol{g}_{\boldsymbol{x}_f}\int_{t_0}^{t_f}\boldsymbol{\Phi}(t_f,t)\boldsymbol{f}_{\boldsymbol{u}}\frac{\delta \boldsymbol{u}}{\delta \tau}\mathrm{d}t + (\boldsymbol{g}_{\boldsymbol{x}_f}\boldsymbol{f} + \boldsymbol{g}_{t_f})\Big|_{t_f}\frac{\delta t_f}{\delta \tau}\right)$$

With Eqs. (28) and (29), and because now Eq. (27) holds, then

$$\frac{\delta J}{\delta \tau} = -k_{t_f}\left(\varphi_t + \varphi_{\boldsymbol{x}}^{\text{T}}\boldsymbol{f} + L + \pi^{\text{T}}(\boldsymbol{g}_{\boldsymbol{x}_f}\boldsymbol{f} + \boldsymbol{g}_{t_f})\right)^2\Big|_{t_f} - \int_{t_0}^{t_f}\left(\boldsymbol{p}_{\boldsymbol{u}} + \boldsymbol{f}_{\boldsymbol{u}}^{\text{T}}\boldsymbol{\Phi}^{\text{T}}(t_f,t)\boldsymbol{g}_{\boldsymbol{x}_f}^{\text{T}}\pi\right)^{\text{T}}\boldsymbol{K}\left(\boldsymbol{p}_{\boldsymbol{u}} + \boldsymbol{f}_{\boldsymbol{u}}^{\text{T}}\boldsymbol{\Phi}^{\text{T}}(t_f,t)\boldsymbol{g}_{\boldsymbol{x}_f}^{\text{T}}\pi\right)\mathrm{d}t \quad (48)$$

This means $\dfrac{\delta J}{\delta \tau} \leq 0$ and $\dfrac{\delta J}{\delta \tau} = 0$ occurs only when Eqs. (33) and (34) hold.

For the optimal value of $\pi$, since $\boldsymbol{K}$ may be arbitrary right-dimensional positive-definite matrix and $k_{t_f}$ may be arbitrary positive constant, we consider three case i) $\boldsymbol{K} = \boldsymbol{1}$, $k_{t_f} = 1$, ii) $\boldsymbol{K} = 2\boldsymbol{1}$, $k_{t_f} = 1$, and iii) $\boldsymbol{K} = \boldsymbol{1}$, $k_{t_f} = 2$, where $\boldsymbol{1}$ is the $m \times m$ dimensional identity matrix. By comparing the three cases of substituting the specific values into Eq. (30), we may obtain Eq. (35), which is irrelevant to the specific value of $\boldsymbol{K}$ and $k_{t_f}$. ∎

Regarding the linear equation (30), assuming that the control satisfies the controllability requirement [19], then the solution is guaranteed. When $\boldsymbol{M}$ is invertible, the parameter $\pi$ may be calculated as

$$\pi = -\boldsymbol{M}^{-1}\boldsymbol{r} \quad (49)$$

*C. Equivalence to the classic optimality conditions*

Actually, Eqs. (33) and (34) are the first-order costate-free optimality conditions for Problem 2. We will show that they are equivalent to the traditional ones with costates [17]. By the adjoining method [14], we may constructed the functional as

$$\bar{J} = \varphi(\boldsymbol{x}(t_f), t_f) + \bar{\boldsymbol{\pi}}^{\mathrm{T}} \boldsymbol{g}(\boldsymbol{x}(t_f), t) + \int_{t_0}^{t_f} \left( L + \boldsymbol{\lambda}^{\mathrm{T}} (\boldsymbol{f} - \dot{\boldsymbol{x}}) \right) \mathrm{d}t \tag{50}$$

where $\boldsymbol{\lambda} \in \mathbb{R}^n$ is the costate variable vector and $\bar{\boldsymbol{\pi}} \in \mathbb{R}^q$ is the Lagrange multiplier parameter. Then the first-order variation may be derived as

$$\delta \bar{J} = \left( \varphi_{t_f} + \bar{\boldsymbol{\pi}}^{\mathrm{T}} \boldsymbol{g}_{t_f} + H \right) \Big|_{t_f} \delta t_f + \left( \boldsymbol{\lambda}(t_f) - \varphi_{\boldsymbol{x}}(t_f) - \boldsymbol{g}_{\boldsymbol{x}_f}^{\mathrm{T}} \bar{\boldsymbol{\pi}} \right) \delta \boldsymbol{x}(t_f) + \int_{t_0}^{t_f} \left( (H_{\boldsymbol{\lambda}} - \dot{\boldsymbol{x}})^{\mathrm{T}} \delta \boldsymbol{\lambda} + (H_{\boldsymbol{x}} + \dot{\boldsymbol{\lambda}})^{\mathrm{T}} \delta \boldsymbol{x} + H_{\boldsymbol{u}}^{\mathrm{T}} \delta \boldsymbol{u} \right) \mathrm{d}t \tag{51}$$

where $H = L + \boldsymbol{\lambda}^{\mathrm{T}} \boldsymbol{f}$ is the Hamiltonian. Through $\delta \bar{J} = 0$, we have

$$\dot{\boldsymbol{\lambda}} + H_{\boldsymbol{x}} = \dot{\boldsymbol{\lambda}} + L_{\boldsymbol{x}} + \boldsymbol{f}_{\boldsymbol{x}}^{\mathrm{T}} \boldsymbol{\lambda} = \boldsymbol{0} \tag{52}$$

$$H_{\boldsymbol{u}} = L_{\boldsymbol{u}} + \boldsymbol{f}_{\boldsymbol{u}}^{\mathrm{T}} \boldsymbol{\lambda} = \boldsymbol{0} \tag{53}$$

and the transversality conditions

$$H(t_f) + \varphi_{t_f} + \bar{\boldsymbol{\pi}}^{\mathrm{T}} \boldsymbol{g}_{t_f} = 0 \tag{54}$$

$$\boldsymbol{\lambda}(t_f) - \varphi_{\boldsymbol{x}}(t_f) - \boldsymbol{g}_{\boldsymbol{x}_f}^{\mathrm{T}} \bar{\boldsymbol{\pi}} = \boldsymbol{0} \tag{55}$$

**Theorem 2:** For Problem 2, the optimality conditions given by Eqs. (33) and (34) are equivalent to the optimality conditions given by (52)-(54).

Proof: Define a quantity $\boldsymbol{\gamma}(t)$ as

$$\boldsymbol{\gamma}(t) = \varphi_{\boldsymbol{x}}(t) + \boldsymbol{\Phi}_o^{\mathrm{T}}(t_f, t) \boldsymbol{g}_{\boldsymbol{x}_f}^{\mathrm{T}} \boldsymbol{\pi} + \int_t^{t_f} \boldsymbol{\Phi}_o^{\mathrm{T}}(\sigma, t) \left( L_{\boldsymbol{x}}(\sigma) + \varphi_{t\boldsymbol{x}}(\sigma) + \varphi_{\boldsymbol{xx}}^{\mathrm{T}}(\sigma) \boldsymbol{f}(\sigma) + \boldsymbol{f}_{\boldsymbol{x}}(\sigma)^{\mathrm{T}} \varphi_{\boldsymbol{x}}(\sigma) \right) \mathrm{d}\sigma \tag{56}$$

Obviously, when $t = t_f$, there is

$$\boldsymbol{\gamma}(t_f) = \varphi_{\boldsymbol{x}}(t_f) + \boldsymbol{g}_{\boldsymbol{x}_f}^{\mathrm{T}} \boldsymbol{\pi} \tag{57}$$

Then Eq. (33) is simplified as

$$L_{\boldsymbol{u}} + \boldsymbol{f}_{\boldsymbol{u}}^{\mathrm{T}} \boldsymbol{\gamma} = \boldsymbol{0} \tag{58}$$

Differentiate $\boldsymbol{\gamma}(t)$ with respect to $t$. In the process, we will use the Leibniz rule [20]

$$\frac{\mathrm{d}}{\mathrm{d}t} \left( \int_{b(t)}^{a(t)} h(\sigma, t) \mathrm{d}\sigma \right) = h(a(t), t) \frac{\mathrm{d}}{\mathrm{d}t} a(t) - h(b(t), t) \frac{\mathrm{d}}{\mathrm{d}t} b(t) + \int_{b(t)}^{a(t)} h_t(\sigma, t) \mathrm{d}\sigma \tag{59}$$

and the property of $\boldsymbol{\Phi}_o(\sigma, t)$ [18]

$$\frac{\partial \boldsymbol{\Phi}_o(\sigma, t)}{\partial t} = -\boldsymbol{\Phi}_o(\sigma, t) \boldsymbol{f}_{\boldsymbol{x}}(t) \tag{60}$$

$$\boldsymbol{\Phi}_o(t, t) = \boldsymbol{I} \tag{61}$$

where $\boldsymbol{I}$ is the $n \times n$ dimensional identity matrix. Then we have

$$\begin{aligned}
\frac{\mathrm{d}}{\mathrm{d}t} \boldsymbol{\gamma}(t) &= \varphi_{t\boldsymbol{x}} + \varphi_{\boldsymbol{xx}}^{\mathrm{T}} \boldsymbol{f} - \boldsymbol{f}_{\boldsymbol{x}}^{\mathrm{T}} \boldsymbol{\Phi}_o^{\mathrm{T}}(t_f, t) \boldsymbol{g}_{\boldsymbol{x}_f}^{\mathrm{T}} \boldsymbol{\pi} - \left( L_{\boldsymbol{x}} + \varphi_{t\boldsymbol{x}} + \varphi_{\boldsymbol{xx}}^{\mathrm{T}} \boldsymbol{f} + \boldsymbol{f}_{\boldsymbol{x}}^{\mathrm{T}} \varphi_{\boldsymbol{x}} \right) - \boldsymbol{f}_{\boldsymbol{x}}^{\mathrm{T}} \int_t^{t_f} \boldsymbol{\Phi}_o^{\mathrm{T}}(\sigma, t) \left( L_{\boldsymbol{x}}(\sigma) + \varphi_{t\boldsymbol{x}}(\sigma) + \varphi_{\boldsymbol{xx}}^{\mathrm{T}}(\sigma) \boldsymbol{f}(\sigma) + \boldsymbol{f}_{\boldsymbol{x}}(\sigma)^{\mathrm{T}} \varphi_{\boldsymbol{x}}(\sigma) \right) \mathrm{d}\sigma \\
&= -L_{\boldsymbol{x}} - \boldsymbol{f}_{\boldsymbol{x}}^{\mathrm{T}} \left( \varphi_{\boldsymbol{x}}(t) + \boldsymbol{\Phi}_o^{\mathrm{T}}(t_f, t) \boldsymbol{g}_{\boldsymbol{x}_f}^{\mathrm{T}} \boldsymbol{\pi} + \int_t^{t_f} \boldsymbol{\Phi}_o^{\mathrm{T}}(\sigma, t) \left( L_{\boldsymbol{x}}(\sigma) + \varphi_{t\boldsymbol{x}}(\sigma) + \varphi_{\boldsymbol{xx}}^{\mathrm{T}}(\sigma) \boldsymbol{f}(\sigma) + \boldsymbol{f}_{\boldsymbol{x}}(\sigma)^{\mathrm{T}} \varphi_{\boldsymbol{x}}(\sigma) \right) \mathrm{d}\sigma \right) \\
&= -L_{\boldsymbol{x}} - \boldsymbol{f}_{\boldsymbol{x}}^{\mathrm{T}} \boldsymbol{\gamma}(t)
\end{aligned} \tag{62}$$

This means $\boldsymbol{\gamma}(t)$ conforms to the same dynamics as the costates $\boldsymbol{\lambda}(t)$ in Eq. (52). Furthermore, use Eqs. (47) and (57), we have

$$\delta J = \left( L + \varphi_t + \boldsymbol{\pi}^{\mathrm{T}} \boldsymbol{g}_{t_f} + \boldsymbol{\gamma}^{\mathrm{T}} \boldsymbol{f} \right)\bigg|_{t_f} \delta t_f + \int_{t_0}^{t_f} \left( \boldsymbol{p_u} + \boldsymbol{f_u}^{\mathrm{T}} \boldsymbol{\Phi}^{\mathrm{T}}(t_f, t) \boldsymbol{g}_{x_f}^{\mathrm{T}} \boldsymbol{\pi} \right)^{\mathrm{T}} \delta \boldsymbol{u} \, \mathrm{d} t \qquad (63)$$

which hold in the feasible solution domain $\mathbb{D}_o$. Compare Eq. (63) with Eq. (51), because $\delta t_f$ may be arbitrary small, to achieve the extremal condition, Eqs. (34) and (54) should be same, i.e.

$$\left( L + \varphi_{t_f} + \bar{\boldsymbol{\pi}}^{\mathrm{T}} \boldsymbol{g}_{t_f} + \boldsymbol{\lambda}^{\mathrm{T}} \boldsymbol{f} \right)\bigg|_{t_f} = \left( L + \varphi_t + \boldsymbol{\pi}^{\mathrm{T}} \boldsymbol{g}_{t_f} + \boldsymbol{\gamma}^{\mathrm{T}} \boldsymbol{f} \right)\bigg|_{t_f} \qquad (64)$$

Since Eq. (64) is generally hold for arbitrary $\boldsymbol{g}$ and $\boldsymbol{f}$, we can conclude that

$$\boldsymbol{\pi} = \bar{\boldsymbol{\pi}} \qquad (65)$$

$$\boldsymbol{\gamma}(t_f) = \boldsymbol{\lambda}(t_f) \qquad (66)$$

which implies the sameness of Eq. (57) and Eq. (55). With Eq. (62), and because $\boldsymbol{\gamma}(t)$ also conforms to the same boundary conditions as the costates $\boldsymbol{\lambda}(t)$, the relation that $\boldsymbol{\gamma}(t) = \boldsymbol{\lambda}(t)$ is established. Therefore Eqs. (58) and (53) are identical. ∎

Again, by investigating the optimality condition (33), it is found that the optimal control are related to the future state, and thus the optimal feedback control law in the analytic form does not exists for the OCPs. From Theorem 2, we actually have got the explicit analytic expression of the costates $\boldsymbol{\lambda}$ and the Lagrange multipliers $\bar{\boldsymbol{\pi}}$ for the classic treatment in Eq. (50), which formerly can only be obtained numerically by solving the BVP. After the proof of Theorem 2, now the variables evolving direction using the VEM is easy to determine.

**Theorem 3:** Solving the IVP with respect to $\tau$, defined by the variation dynamic evolution equations (8), (28) and (29) from a feasible initial solution, when $\tau \to +\infty$, $(\boldsymbol{x}, \boldsymbol{u})$ will satisfy the optimality conditions of Problem 2.

Proof: By Lemma 2 and with Eq. (18) as the Lyapunov functional, we may claim that the minimum solution of Peoblem 2 is an asymptotically stable solution within the feasibility domain $\mathbb{D}_o$ for the infinite-dimensional dynamics governed by Eqs. (8), (28) and (29). From a feasible initial solution, any evolution under these dynamics maintains the feasibility of the variables, and they also guarantee $\frac{\delta J}{\delta \tau} \leq 0$. The functional $J$ will decrease until $\frac{\delta J}{\delta \tau} = 0$, which occurs when $\tau \to +\infty$ due to the asymptotical approach. When $\frac{\delta J}{\delta \tau} = 0$, this determines the optimal conditions, namely, Eqs. (33) and (34). ∎

Presume that we already have a feasible initial solution $\tilde{\boldsymbol{x}}(t)$, $\tilde{\boldsymbol{u}}(t)$ and $\tilde{t}_f$ that satisfy Eqs. (19), (20) and (21), Theorem 3 guarantees the infinite-dimensional variation dynamic evolution equations (8), (28) and (29) may be used to obtain the optimal solution that minimizes Eq. (18).

*D. Formulation of EPDE*

Use the partial differential operator "$\partial$" and the differential operator "d" to reformulate the variation dynamic evolution equations, we may get the EPDE and EDE as

$$\frac{\partial}{\partial \tau} \begin{bmatrix} \boldsymbol{x} \\ \boldsymbol{u} \end{bmatrix} = \begin{bmatrix} \int_{t_0}^{t} \boldsymbol{\Phi}_o(t,s) \boldsymbol{f_u}(s) \dfrac{\partial \boldsymbol{u}}{\partial \tau}(s) \, \mathrm{d} s \\ -\boldsymbol{K} \left( \boldsymbol{p_u} + \boldsymbol{f_u}^{\mathrm{T}} \boldsymbol{\Phi}^{\mathrm{T}}(t_f, t) \boldsymbol{g}_{x_f}^{\mathrm{T}} \boldsymbol{\pi} \right) \end{bmatrix} \qquad (67)$$

$$\frac{\mathrm{d}t_f}{\mathrm{d}\tau} = -k_{t_f}\left(L + \varphi_t + \varphi_x^\mathrm{T} f + \pi^\mathrm{T}(g_{x_f} f + g_{t_f})\right)\Big|_{t_f} \tag{68}$$

Put into this perspective, the definite conditions are $t_f\big|_{\tau=0} = \tilde{t}_f$ and $\begin{bmatrix} x(t,\tau) \\ u(t,\tau) \end{bmatrix}\bigg|_{\tau=0} = \begin{bmatrix} \tilde{x}(t) \\ \tilde{u}(t) \end{bmatrix}$, where $\tilde{x}(t)$ and $\tilde{u}(t)$ are the initial feasible solution. Eqs. (67) and (68) realize the anticipated variable evolving along the variation time $\tau$ as depicted in Fig. 1. The initial conditions of $x(t,\tau)$ and $u(t,\tau)$ at $\tau = 0$ belong to the feasible solution domain and their value at $\tau = +\infty$ represents the optimal solution of the OCP. The right part of the EPDE (67) is also only a vector function of time $t$. Thus we may apply the semi-discrete method to discretize it along the normal time dimension and further use ODE integration methods to get the numerical solution.

As already pointed out previously that Problem 2 contains Problem 1, thus the results developed in this paper are of more general meaning. See the evolution equations, when the terminal states are free, i.e., $g = 0$, they are degraded to the case in Ref. [13]. Moreover, this paper considers the OCPs defined with free terminal time, yet the results obtained are also applicable to the simpler case with fixed terminal time. For those OCPs, the equation regarding the terminal time $t_f$ is not necessary anymore, while the evolution equations regarding state variable $x$ and control variable $u$ are still similar with $M$, $r$ in Eq. (30) simplified as

$$M = g_{x_f}\left(\int_{t_0}^{t_f} \Phi(t_f,t) f_u K f_u^\mathrm{T} \Phi^\mathrm{T}(t_f,t)\,\mathrm{d}t\right) g_{x_f}^\mathrm{T} \tag{69}$$

$$r = g_{x_f}\left(\int_{t_0}^{t_f} \Phi(t_f,t) f_u K p_u\,\mathrm{d}t\right) \tag{70}$$

### IV. ILLUSTRATIVE EXAMPLES

First a linear example taken from Xie [21] is considered.

**Example 1**: Consider the following dynamic system

$$\dot{x} = Ax + bu$$

where $x = \begin{bmatrix} x_1 \\ x_2 \end{bmatrix}$, $A = \begin{bmatrix} 0 & 1 \\ 0 & 0 \end{bmatrix}$, $b = \begin{bmatrix} 0 \\ 1 \end{bmatrix}$. Find the solution that minimizes the performance index

$$J = \frac{1}{2}\int_{t_0}^{t_f} u^2\,\mathrm{d}t$$

with the boundary conditions

$$x(t_0) = \begin{bmatrix} 1 \\ 1 \end{bmatrix}, \; x(t_f) = \begin{bmatrix} 0 \\ 0 \end{bmatrix}$$

where the initial time $t_0 = 0$ and the terminal time $t_f = 2$ are fixed.

In solving this example using the VEM, the EPDE derived is

$$\frac{\partial}{\partial \tau}\begin{bmatrix} x \\ u \end{bmatrix} = \begin{bmatrix} \int_{t_0}^{t} e^{A(t-s)} b \frac{\partial u}{\partial \tau}(s)\,\mathrm{d}s \\ -K\left\{u + b^\mathrm{T}\left(e^{A(t_f-t)}\right)^\mathrm{T}\pi\right\} \end{bmatrix}$$

In particular, for this problem, $\pi$ is a constant that may be calculated as

$$\boldsymbol{\pi} = -\boldsymbol{M}^{-1}\boldsymbol{r} = \begin{bmatrix} 3.0 \\ -2.5 \end{bmatrix}$$

where $\boldsymbol{M} = K\int_{t_0}^{t_f} e^{A(t_f-t)}\boldsymbol{b}\boldsymbol{b}^{\mathrm{T}}\left(e^{A(t_f-t)}\right)^{\mathrm{T}}\mathrm{d}t$ and $\boldsymbol{r} = K\int_{t_0}^{t_f} e^{A(t_f-t)}\boldsymbol{b}u\,\mathrm{d}t = -Ke^{A(t_f-t_0)}\boldsymbol{x}_0$. The one-dimensional matrix $K$ was set as $K = 0.1$. The definite conditions of the EPDE, i.e., the initial guess of the states $\tilde{\boldsymbol{x}}(t)$ and the control $\tilde{u}(t)$, were obtained by numerical integration. To achieve the feasibility, we designed the following control law as

$$\tilde{u}(t) = -\omega_n^2 x_1(t) - 2\omega_n \xi x_2(t)$$

where the damp parameter is $\xi = 0.707$ and the frequency parameter is set to be time-varying as $\omega_n = 5t$. Using the semi-discrete method, the time horizon $[t_0, t_f]$ was discretized uniformly with 41 points. Thus, a dynamic system with 123 states was obtained and the OCP was transformed to a finite-dimensional IVP. The ODE integrator "ode45" in Matlab, with default relative error tolerance $1\times 10^{-3}$ and default absolute error tolerance $1\times 10^{-6}$, was employed to solve the IVP. For comparison, the analytic solution by solving the BVP is presented.

$$\begin{aligned}
\hat{x}_1 &= 0.5t^3 - 1.75t^2 + t + 1 \\
\hat{x}_2 &= 1.5t^2 - 3.5t + 1 \\
\hat{\lambda}_1 &= 3 \\
\hat{\lambda}_2 &= -3t + 3.5 \\
\hat{u} &= 3t - 3.5
\end{aligned}$$

Figs. 2, 3 and 4 show the evolving process of $x_1(t)$, $x_2(t)$ and $u(t)$ solutions to the optimal, respectively. At $\tau = 300$s, the numerical solutions are indistinguishable from the optimal, and this shows the effectiveness of the VEM. Fig. 5 plots the profile of performance index value against the variation time. It declines rapidly at first and almost reaches the minimum when $\tau = 40$s. Then it keeps approaching the analytic minimum of 3.25 monotonously. In addition, from Eq. (56), we may compute that

$$\boldsymbol{\gamma}(t) = \boldsymbol{\Phi}_o^{\mathrm{T}}(t_f, t)\boldsymbol{\pi} = \left(e^{A(t_f-t)}\right)^{\mathrm{T}}\boldsymbol{\pi} = \begin{bmatrix} 1 & 0 \\ t_f - t & 1 \end{bmatrix}\begin{bmatrix} 3.0 \\ -2.5 \end{bmatrix} = \begin{bmatrix} 3 \\ -3t + 3.5 \end{bmatrix}$$

It is exactly identical to $\begin{bmatrix} \hat{\lambda}_1 \\ \hat{\lambda}_2 \end{bmatrix}$, as proved in Theorem 2.

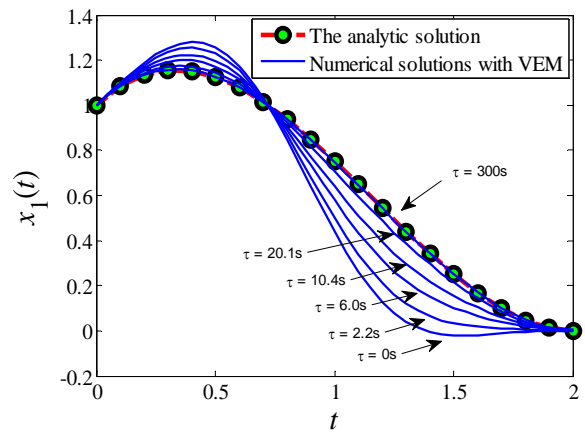

Fig. 2 The evolution of numerical solutions of $x_1$ to the optimal solution.

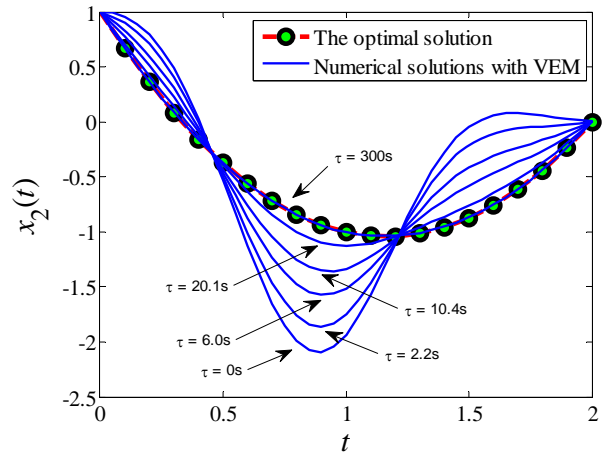

Fig. 3 The evolution of numerical solutions of $x_2$ to the optimal solution.

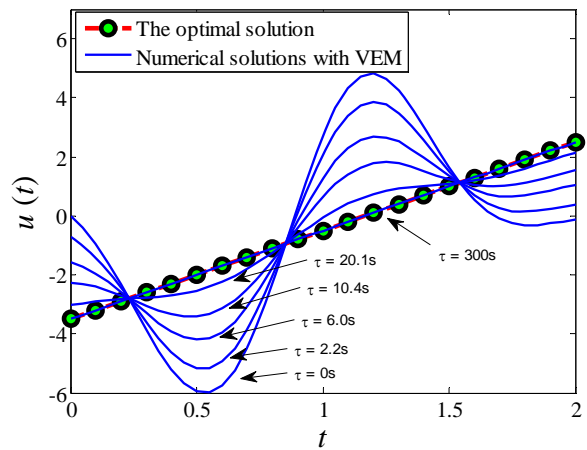

Fig. 4 The evolution of numerical solutions of $u$ to the optimal solution.

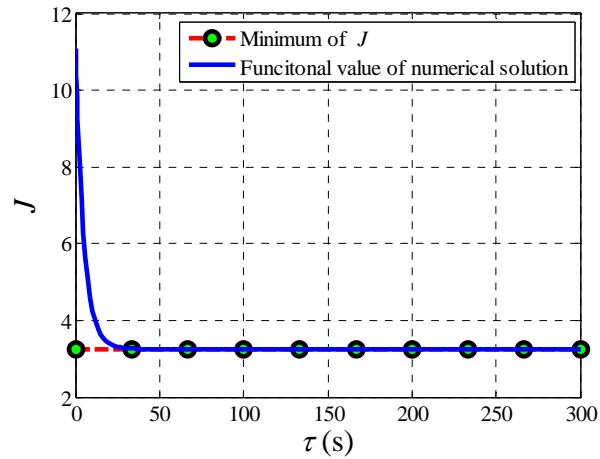

Fig. 5 The approach to the minimum of performance index.

Now we consider a nonlinear example with free terminal time $t_f$, the Brachistochrone problem [22], which describes the motion curve of the fastest descending.

**Example 2**: Consider the following dynamic system

$$\dot{x} = f(x, u)$$

where $x = \begin{bmatrix} x \\ y \\ V \end{bmatrix}$, $f = \begin{bmatrix} V\sin(u) \\ -V\cos(u) \\ g\cos(u) \end{bmatrix}$, $g = 10$ is the gravity constant. Find the solution that minimizes the performance index

$$J = t_f$$

with the boundary conditions

$$\begin{bmatrix} x \\ y \\ V \end{bmatrix}\bigg|_{t_0=0} = \begin{bmatrix} 0 \\ 0 \\ 0 \end{bmatrix}, \begin{bmatrix} x \\ y \end{bmatrix}\bigg|_{t_f} = \begin{bmatrix} 2 \\ -2 \end{bmatrix}$$

This example has fixed terminal position boundary conditions and free terminal velocity $V(t_f)$. Thus there is

$$g_{x_f} = \begin{bmatrix} 1 & 0 & 0 \\ 0 & 1 & 0 \end{bmatrix}$$

In the specific form of the EPDE (67) and the EDE (68), the parameters $K$ and $k_{t_f}$ were set to be 0.1 and 0.05, respectively. The definite conditions, i.e., $\begin{bmatrix} x(t,\tau) \\ u(t,\tau) \\ t_f(\tau) \end{bmatrix}\bigg|_{\tau=0}$, were obtained from a physical motion along a straight line that connects the initial position to the terminal position, that is

$$\tilde{t}_f = \sqrt{0.8} \qquad \tilde{u} = \frac{\pi}{4}$$
$$\tilde{x} = 2.5t^2 \qquad \tilde{y} = -2.5t^2 \qquad \tilde{V} = 5\sqrt{2}t$$

We also discretized the time horizon $[t_0, t_f]$ uniformly, with 101 points. Thus, a large IVP with 405 states (including the terminal time) is obtained. We still employed "ode45" in Matlab for the numerical integration. In the integrator setting, the default relative error tolerance and the absolute error tolerance are $1 \times 10^{-3}$ and $1 \times 10^{-6}$, respectively. For comparison, we computed the optimal solution with GPOPS-II [23], a Radau PS method based OCP solver.

Fig. 6 gives the states curve in the $xy$ coordinate plane, showing that the numerical results starting from a straight line approach the optimal solution over time. The control solutions are plotted in Fig. 7, and the asymptotical approach of the numerical results are demonstrated. In Fig. 8, the terminal time profile against the variation time $\tau$ is plotted. The result of $t_f$ declines rapidly at first and then gradually approaches to the minimum decline time, and it only changes slightly after $\tau = 40$s. At $\tau = 300$s, we compute that $t_f = 0.8168$s from the VEM, very close to the result of 0.8165s from GPOPS-II. Fig. 9 presents the evolution profiles of the Lagrange multipliers $\pi$. They also approach the optimal value of $\begin{bmatrix} -0.1477 \\ 0.0564 \end{bmatrix}$ rapidly.

15

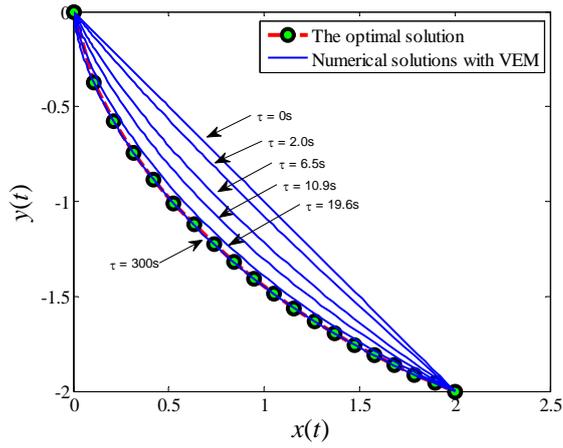

Fig. 6 The evolution of numerical solutions in the $xy$ coordinate plane to the optimal solution.

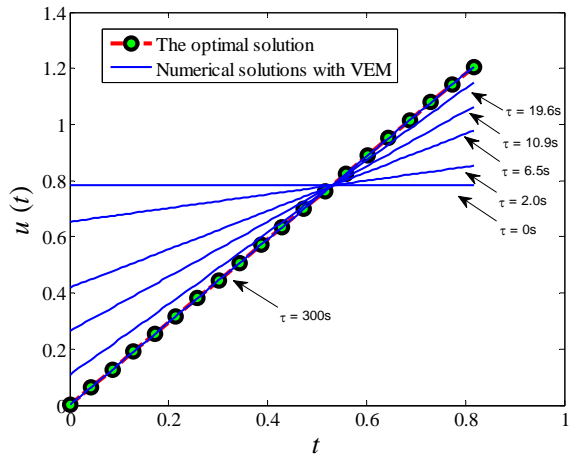

Fig. 7. The evolution of numerical solutions of $u$ to the optimal solution.

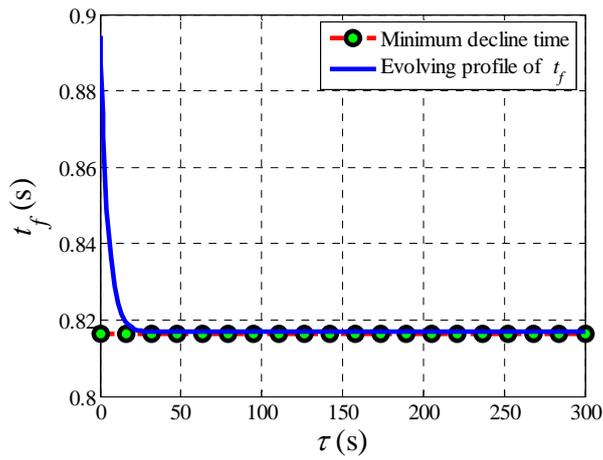

Fig. 8 The evolution profile of $t_f$ to the minimum decline time.

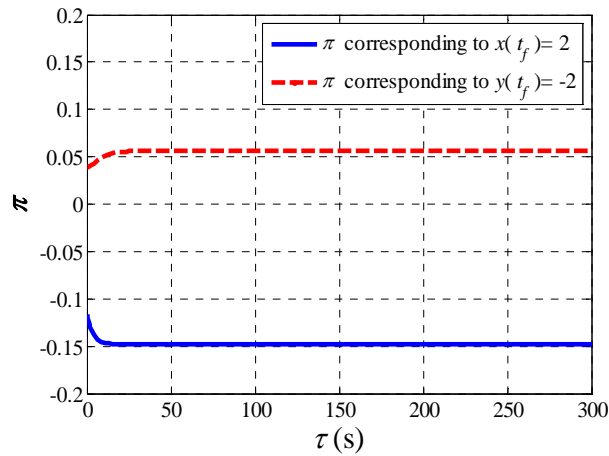

Fig. 9 The evolution profiles of the Lagrange multipliers $\boldsymbol{\pi}$.

## V. Conclusion

This paper further developes the compact Variation Evolving Method (VEM) to address the computation of Optimal Control Problems (OCPs) with terminal constraint. A set of more general evolution equations is derived, and the costate-free optimality conditions are established for OCPs with prescribed terminal boundary conditions. Especially, from the equivalence proof of the optimality conditions, even if the costates are completely not considered during the derivation, its analytic expressions related to the states and the control variables are obtained. The analytic relations between Lagrange multipliers, which adjoin the terminal constraints, and the original variables are also explicitly got. These results may help us to deepen the understanding towards the optimal control theory. However, currently the proposed method requires the initial solution to be feasible. This is inflexible, especially for the problem with terminal constraint, and further studies will be carried out to address this issue.